\documentclass[a4paper,10pt,twoside]{cpc-hepnp}

\usepackage{multicol}
\usepackage{graphicx}
\usepackage{booktabs}
\usepackage{amssymb,bm,mathrsfs,bbm,amscd}
\usepackage[tbtags]{amsmath}
\usepackage{lastpage}
\usepackage{CJK}

\usepackage{graphicx,color,dcolumn,booktabs,bm}
\usepackage{longtable,lscape,comment,braket}
\usepackage{txfonts}
\usepackage{overpic}
\usepackage{amssymb}
\usepackage{indentfirst}
\usepackage{feynmf}   
\usepackage{slashed}  
\usepackage{cases}
\usepackage{epstopdf}
\usepackage{psfrag}
\usepackage{subfigure}
\usepackage[colorlinks,
            citecolor=blue,
            anchorcolor=red,
            menucolor=red,
            linkcolor=red,
            filecolor=red,
            runcolor=red,
            urlcolor=blue,
            frenchlinks=red]{hyperref}

\begin{document}
\begin{CJK*}{GB}{gbsn}

\fancyhead[c]{\small Chinese Physics C~~~Vol. **, No. * (****)
******} \fancyfoot[C]{\small ******-\thepage}

\footnotetext[0]{Received \today}

\title{Possible strange hidden-charm pentaquarks from  $\Sigma_c^{(*)}\bar{D}_s^*$ and $\Xi^{(',*)}_c\bar{D}^*$ interactions
\thanks{This project is supported by the
National Natural Science Foundation of China under grants No. 11222547, No. 11175073, No. 11675228 and  the Major State
Basic Research Development Program in China under grant 2014CB845405 and the Fundamental Research Funds for the Central Universities. Xiang Liu is also supported by the National Program for Support of Top-notch Young Professionals.}}

\author{%
      Rui Chen (³ÂÈñ)$^{1,3;1)}$\email{chenr15@lzu.edu.cn}%
\quad Jun He (ºÎ¾ü)$^{2,3,4;2)}$\email{jun.he.1979@icloud.com}%
\quad Xiang Liu (ÁõÏè)$^{1,3,4;3)}$ \email{xiangliu@lzu.edu.cn}
}
\maketitle

\address{%
$^1$School of Physical Science and Technology, Lanzhou University, Lanzhou 730000, China\\
$^2$Department of  Physics, Nanjing Normal University,
Nanjing, Jiangsu 210097, China\\
$^3$Research Center for Hadron and CSR Physics, Lanzhou University
and Institute of Modern Physics of CAS, Lanzhou 730000, China\\
$^4$Nuclear Theory Group, Institute of Modern Physics, Chinese Academy of Sciences, Lanzhou 730000, China
}

\begin{abstract}
Using the one-boson-exchange model, we investigate the $\Lambda_c\bar{D}_s^*$, $\Sigma_c\bar{D}_s^*$, $\Sigma_c^*\bar{D}_s^*$, $\Xi_c\bar{D}^*$, $\Xi_c'\bar{D}^*$, and $\Xi_c^*\bar{D}^*$ interactions by considering the one-eta-exchange and/or one-pion-exchange contributions. We further predict the existence of
hidden-charm molecular pentaquarks. Promising candidates for hidden-charm molecular pentaquarks include a $\Xi_c'\bar{D}^*$ state with $0(\frac{1}{2}^-)$ and the $\Xi_c^*\bar{D}^*$ states with $0(\frac{1}{2}^-)$ and $0(\frac{3}{2}^-)$. Experimental searches for these predicted hidden-charm molecular pentaquarks are an interesting future research topic for experiments like LHCb.
\end{abstract}



\footnotetext[0]{\hspace*{-3mm}\raisebox{0.3ex}{$\scriptstyle\copyright$}2013
Chinese Physical Society and the Institute of High Energy Physics
of the Chinese Academy of Sciences and the Institute
of Modern Physics of the Chinese Academy of Sciences and IOP Publishing Ltd}%

\begin{multicols}{2}

\section{Introduction}

As a hot research issue, studying exotic hadronic states is attractive for both experimentalists and theorists. In the past 14 years, more and more charmonium-like and bottomonium-like states and states with open-charm and open-bottom quantum numbers have been reported in experiments. This gives us a good chance to identify possible candidates for exotic states (see the comprehensive reviews in Refs. \cite{Chen:2016qju,Chen:2016spr} for recent progress in this field).

There are different configurations of exotic states, which include glueballs, hybrids, molecular states, multiquark states and so on. Among these configurations, the molecular state is very popular to study these newly observed hadronic states. In  observed hadronic matter, the deuteron, which is composed of a neutron and a proton, has been confirmed to be the typical hadronic molecular state existing in nature. The measured binding energy of deuteron is $E=-2.224575$ MeV \cite{Machleidt:2000ge,Wiringa:1994wb}. To quantitatively depict a neutron interacting with a proton to form a deuteron, theorists have focused on the nuclear force and developed  corresponding phenomenological models like the one-boson-exchange (OBE) model. Here, pions, sigmas and rho/omega particles contribute to the nuclear force at long, medium and short distances, respectively. Since 2003, the observed charmonium-like and bottomonium-like states have stimulated extensive interest in applying the OBE model to investigate the newly observed $X(3872)$ \cite{Tornqvist:1993ng,Tornqvist:1993vu,Liu:2008fh,Thomas:2008ja,Lee:2009hy,Li:2012cs,Sun:2012zzd,Zhao:2014gqa}, $Y(3930)$/$Y(4140)$ \cite{Liu:2009ei,Liu:2008tn,Ding:2009vd} and $Z_b(10610)$/$Z_b(10650)$ \cite{Liu:2008fh,Sun:2011uh,Dias:2014pva}. In addition, the interactions of two hadrons have been explored under the OBE model and more hadronic molecular states predicted \cite{Liu:2008du,Hu:2010fg,Shen:2010ky,Lee:2011rka,Li:2012bt,Li:2012ss,Zhao:2015mga,Yang:2011wz,Chen:2015loa,Chen:2016heh,Chen:2014mwa,Chen:2016ypj,Chen:2015add}.

In 2015, $P_c(4380)$ and $P_c(4450)$ were reported by the LHCb Collaboration~\cite{Aaij:2015tga}. Before the observation of these two $P_c$ states, theorists once predicted the existence of molecular hidden-charm pentaquarks \cite{Wu:2010jy,Wu:2010vk,Hofmann:2005sw,Yang:2011wz,Wang:2011rga,Yuan:2012wz,Karliner:2015ina}. Later, combining with the released experimental information, theorists analyzed the properties of the two $P_c$ states by different approaches \cite{Chen:2015loa,Chen:2016heh,Chen:2015moa,Roca:2015dva,Mironov:2015ica,He:2015cea,Meissner:2015mza,Burns:2015dwa,Huang:2015uda,Maiani:2015vwa,Anisovich:2015cia,
Li:2015gta,Ghosh:2015ksa,Wang:2015epa,Anisovich:2015zqa,Lebed:2015tna,Zhu:2015bba,Guo:2015umn,Liu:2015fea,Mikhasenko:2015vca,
Scoccola:2015nia} (see Ref. \cite{Chen:2016qju} and references therein for details).

Under the molecular state assignment to $P_c(4380)$ and $P_c(4450)$ \cite{Chen:2015loa,Chen:2016heh,Chen:2015moa,Roca:2015dva,Mironov:2015ica,He:2015cea,Meissner:2015mza,
Burns:2015dwa,Huang:2015uda}, the interactions of charmed baryons with anti-charmed mesons were studied. If $P_c(4380)$ and $P_c(4450)$ are molecular pentaquarks composed of a charmed baryon and an anti-charmed meson, we have reason to believe that their partners should exist.
For example, in Ref.~\cite{Chen:2016heh}, possible charm-strange molecular pentaquarks were studied by the OBE model. The authors suggested $\Lambda_b^0\to\bar{D}^0D^0\Lambda^0$ as an appropriate channel to search for two predicted pentaquarks $P_{cs}(3340)$ and $P_{cs}(3400)$, which correspond to the $\Sigma_c\bar{K}^*$ configuration with $I(J^P)=\frac{1}{2}(\frac{3}{2}^-)$ and the $\Sigma_c^*\bar{K}^*$ configuration with $\frac{1}{2}(\frac{5}{2}^-)$, respectively.

Along this line, in this work we focus on other hidden-charm molecular pentaquarks with a strange quark, which have the concrete quark component $[c\bar{c}sqq]$. These molecular pentaquarks, which are also called strange hidden-charm molecular pentaquarks in this work, are closely related to the $\Lambda_c\bar{D}_s^*/\Sigma_c^{(*)}\bar{D}_s^*/\Xi^{(\prime,*)}_c\bar{D}^*$ interactions. There are some previous theoretical studies of strange hidden-charm molecular pentaquarks~\cite{Wu:2010jy,Wu:2010vk,Hofmann:2005sw}.
Very recently, Karliner and Rosner ~\cite{Karliner:2016ith} proposed that $\Lambda_b\rightarrow J/\psi\Lambda(\pi^+\pi^-/\eta)$ is a promising channel to find a possible strange hidden-charm molecular pentaquark composed of $\Lambda_c$ and $\bar{D}_s^*$.

Different from the former studies of strange hidden-charm molecular pentaquark in Refs. \cite{Wu:2010jy,Wu:2010vk,Hofmann:2005sw,Karliner:2016ith}, in this work we carry out a comprehensive investigation of the $\Lambda_c\bar{D}_s^*/\Sigma_c^{(*)}\bar{D}_s^*/\Xi^{(\prime,*)}_c\bar{D}^*$ interactions under the OBE model, by which we further predict the strange hidden-charm molecular pentaquarks. Here, two kinds of molecular configurations will be taken into consideration: molecular systems composed of a charmed baryon and an anti-charmed-strange meson (molecular pentaquarks with components $\Lambda_c\bar{D}_s^*$, $\Sigma_c\bar{D}_s^*$, and $\Sigma_c^*\bar{D}_s^*$), and systems composed of a charmed-strange baryon and an anti-charmed meson (molecular pentaquarks with components $\Xi_c\bar{D}^*$, $\Xi_c'\bar{D}^*$, and $\Xi_c^*\bar{D}^*$). In the following section, we will give a detailed illustration of deducing the effective potentials involved in the study of strange hidden-charm molecular pentaquarks.
We hope that the present work may provide valuable information about strange hidden-charm molecular pentaquarks, which will be helpful for further experimental searches for them.

The paper is organized as follows. We present the deduction of the effective potentials in Section~\ref{sec2}. In Section~\ref{sec3}, the corresponding numerical results for the strange hidden-charm pentaquarks are given. A summary is then given in Section~\ref{sec4}.

\section{Effective potentials related to the $\Lambda_c\bar{D}_s^*/\Sigma_c^{(*)}\bar{D}_s^*/\Xi^{(\prime,*)}_c\bar{D}^*$ systems}\label{sec2}

First, we need to illustrate the details of deducing the effective potential. For the $\Lambda_c\bar{D}_s^*/\Sigma_c^{(*)}\bar{D}_s^*/\Xi^{(\prime,*)}_c\bar{D}^*$ systems, their total wave functions are constructed by including color, flavor, spin-orbit, and spatial wave functions. For colorless molecular states, the color wave function is simply taken as $1$. In addition, we adopt the notation $|{}^{2S+1}L_J\rangle$ to define the spin-orbit wave function. The total angular momentum $J$ can be $\frac{1}{2}$ and $\frac{3}{2}$ for the $\Lambda_c\bar{D}_s^*$, $\Sigma_c\bar{D}_s^*$, $\Xi_c'\bar{D}^*$ and $\Xi_c\bar{D}^*$ systems, and $\frac{1}{2}$, $\frac{3}{2}$ and $\frac{5}{2}$ for the $\Sigma_c^*\bar{D}_s^*$ and $\Xi_c^*\bar{D}^*$ systems. The spin-orbit wave function $|{}^{2S+1}L_J\rangle$ can be explicitly expressed as
\begin{eqnarray}
\begin{array}{ccccc}
J=\frac{1}{2}:    &|{}^2\mathbb{S}_{\frac{1}{2}}\rangle,   &|{}^4\mathbb{D}_{\frac{1}{2}}\rangle;\\
J=\frac{3}{2}:    &|{}^4\mathbb{S}_{\frac{3}{2}}\rangle,   &|{}^2\mathbb{D}_{\frac{3}{2}}\rangle,    &|{}^4\mathbb{D}_{\frac{3}{2}}\rangle;\\
J=\frac{5}{2}:    &|{}^6\mathbb{S}_{\frac{5}{2}}\rangle,   &|{}^2\mathbb{D}_{\frac{5}{2}}\rangle,    &|{}^4\mathbb{D}_{\frac{5}{2}}\rangle,    &|{}^6\mathbb{D}_{\frac{5}{2}}\rangle,
\end{array}
\end{eqnarray}
where $\mathbb{S}$ and $\mathbb{D}$ denote the corresponding systems with orbit angular momentum $L=0$ and $L=2$, respectively. We need to specify that in our calculation we consider the mixing of $S$-wave and $D$-wave, which is the lesson learned from deuteron studies. Here, S-D mixing contributes to the tensor force, which is crucial to form the shallow deuteron.

The explicit expressions for the spin-orbit wave function are categorized into two typical groups by the spin $S_B$ of the baryon in the system, i.e.,
\begin{eqnarray}
\left|{}^{2S+1}L_{J}\right\rangle_{S_B=\frac{1}{2}} &=& \sum_{m,m'}^{m_S,m_L}C^{S,m_S}_{\frac{1}{2}m,1m'}C^{J,M}_{Sm_S,Lm_L}
          \chi_{\frac{1}{2}m}\epsilon^{m'}|Y_{L,m_L}\rangle,\label{fun1}\\
\left|{}^{2S+1}L_{J}\right\rangle_{S_B=\frac{3}{2}} &=& \sum_{m,m'}^{m_S,m_L}C^{S,m_S}_{\frac{3}{2}m,1m'}C^{J,M}_{Sm_S,Lm_L}
          \Phi_{\frac{3}{2}m}\epsilon^{m'}|Y_{L,m_L}\rangle,\label{fun2}
\end{eqnarray}
where $Y_{L,m_L}$ is the spherical harmonics function, and the constants $C^{J,M}_{Sm_S,Lm_L}$, $C^{S,m_S}_{\frac{1}{2}m,1m'}$ and $C^{S,m_S}_{\frac{3}{2}m,1m'}$ are the Clebsch-Gordan coefficients. The polarization vectors for the vector meson are defined as $\epsilon_{\pm}^{m}=\mp\frac{1}{\sqrt{2}}\left(\epsilon_x^{m}{\pm}i\epsilon_y^{m}\right)$ and $\epsilon_0^{m}=\epsilon_z^{m}$, which are written explicitly as $\epsilon_{\pm1} = \frac{1}{\sqrt{2}}\left(0,\pm1,i,0\right)$ and $\epsilon_{0} = \left(0,0,0,-1\right)$. The $\chi_{\frac{1}{2}m}$ denotes the spin wave function for baryons ($\Lambda_c$, $\Sigma_c$, or $\Xi_c^{(')}$) with spin $S_B=\frac{1}{2}$. The polarization tensor $\Phi_{\frac{3}{2}m}$ for baryons ($\Sigma_c^*$ or $\Xi_c^*$, ) with spin $S_B=\frac{3}{2}$ has the form $\Phi_{\frac{3}{2}m}=\sum_{m_1,m_2}\langle \frac{1}{2},m_1;1,m_2|\frac{3}{2},m\rangle\chi_{\frac{1}{2},m_1}\epsilon^{m_2}$.

Additionally, the flavor wave function $|I,I_3\rangle$ of these molecular systems, where $I$ and $I_3$ are the isospin and its third component of the systems, respectively, has the form
\begin{eqnarray}\left.\begin{array}{cl}
\Lambda_c\bar{D}_s^*: &|0,0\rangle=|\Lambda_c^+D_s^-\rangle,\\
\Sigma_c^{(*)}\bar{D}_s^*:  &\left\{\begin{array}{l}|1,1\rangle=|\Sigma_c^{(*)++}D_s^{*-}\rangle,\\
                                              |1,0\rangle=|\Sigma_c^{(*)+}D_s^{*-}\rangle,\\
                                              |1,-1\rangle=|\Sigma_c^{(*)0}D_s^{*-}\rangle,
                       \end{array}\right.\\
\Xi_c^{(\prime,*)}\bar{D}^*:       &\left\{\begin{array}{l}|1,1\rangle=|\Xi_c^{(\prime,*)+}\bar{D}^{*0}\rangle,\\
                                              |1,0\rangle=\frac{1}{\sqrt{2}}\left(|\Xi_c^{(\prime,*)+}D^{*-}\rangle+|\Xi_c^{(\prime,*)0}\bar{D}^{*0}\rangle\right),\\
                                              |1,-1\rangle=|\Xi_c^{(\prime,*)0}D^{*-}\rangle,
                       \end{array}\right.\\
                      &|0,0\rangle=\frac{1}{\sqrt{2}}\left(|\Xi_c^{(\prime,*)+}D^{*-}\rangle-|\Xi_c^{(\prime,*)0}\bar{D}^{*0}\rangle\right),
\end{array}\right.
\end{eqnarray}
where $\Sigma_c^{(*)}$ indicates the charmed baryons $\Sigma_c$ and $\Sigma_c^*$, and the charmed-strange baryons $\Xi_c$, $\Xi_c^\prime$ and $\Xi_c^*$, are denoted by $\Xi_c^{(\prime,*)}$.

In the following, we continue to deduce the effective potentials of the $\Lambda_c\bar{D}_s^*/\Sigma_c^{(*)}\bar{D}_s^*/\Xi^{(\prime,*)}_c\bar{D}^*$ systems. In general, the effective potential in momentum space is related to the scattering amplitude, i.e.,
\begin{eqnarray}
\mathcal{V}^{~ab\to cd}(\bf{q}) &=&
   -\frac{\mathcal{M}(ab\to cd)}
   {\sqrt{2m_a2m_b2m_c2m_d}},
\end{eqnarray}
where $\mathcal{M}(ab\to cd)$ denotes the scattering amplitude of a process $ab\to cd$, and $m_{i}$ ($i=a,b,c,d$) is the mass of particle $a/b/c/d$. At the hadronic level, we can write out the expression of $\mathcal{M}(ab\to cd)$ by the effective Lagrangian approach. Then, an effective potential in momentum space $\mathcal{V}_E(\bf{q})$ can be transferred into an effective potential in coordinate space by performing the Fourier transformation
\begin{eqnarray}
\mathcal{V}_E^{~ab\to cd}({r}) &=&
     \int\frac{d^3 \bf{q}}{(2\pi)^3}e^{i\bf{q}\cdot\bf{r}}
     \mathcal{V}_E^{~ab\to cd}({\bf{q}})\mathcal{F}^2(q^2,m_E^2).
\end{eqnarray}
In the above Fourier transformation, the form factor $\mathcal{F}(q^2,m_E^2)$ should be introduced at each interaction vertex to compensate the off-shell effect of the exchanged meson and reflect the inner structure of each interaction vertex. Usually, a monopole form like $\mathcal{F}(q^2,m_E^2) = (\Lambda^2-m_E^2)/(\Lambda^2-q^2)$ is suggested \cite{Tornqvist:1993ng,Tornqvist:1993vu}\footnote{{In Refs. \cite{Tornqvist:1993ng,Tornqvist:1993vu}, T\"ornqvist studied the  one-pion exchange potential contribution to $N{N}$ systems. Especially, the deuteron was discussed with the introduced monopole form factor, where an expression for a spherical pion source was given,  $R={\sqrt{10}\over\Lambda}=\frac{0.624}{\Lambda/[GeV]}$~fm. According to this relation, the $\Lambda$ value was estimated to be $0.8-1.5$ GeV for the $NN$ interaction. For the discussed hidden-charm molecular pentaquarks, a smaller $R$ should be expected, which results in a larger $\Lambda$. Due to this reason, in this work we choose $\Lambda=0.8-5$ GeV to present our numerical results. Although the bound state solution by scanning this wide $\Lambda$ range can be found, we still should be careful to make a definite conclusion of the existence of the corresponding hidden-charm molecular pentaquarks. Thus, in this work we take a stricter criterion, i.e., if the bound state solutions appear when taking $\Lambda=1-1.5$ GeV, the existence of the corresponding molecular state becomes more possible. In Section \ref{sec3}, we give more detailed discussions.}}\footnote{{Indeed we can take other forms for the form factor, such as the dipole form factor $\mathcal{F}(q^2,m_E^2) = (\Lambda^2-m_E^2)^2/(\Lambda^2-q^2)^2$, to regularize the potential. If taking this form of form factor in the calculation, we need to fix the possible range of $\Lambda$ by restudying the $NN$ interaction.}}, where $m_E$ and $q$ denote the mass and four-momentum of the exchanged particle, respectively. In addition, the cutoff $\Lambda$ is a model parameter in our calculation. Later, we will discuss the dependence of the numerical result on $\Lambda$. With the obtained effective potential, we try to find bound state solutions by solving the Schr\"{o}dinger equation. In this way, we can further predict the mass spectrum of the possible molecular states.

When writing out the scattering amplitude, we adopt the effective Lagrangian approach.
Due to both heavy quark symmetry and chiral symmetry \cite{Yan:1992gz,Burdman:1992gh,Wise:1992hn,Casalbuoni:1996pg,Falk:1992cx,Liu:2011xc}, the relevant Lagrangians can be constructed as
\begin{eqnarray}
\mathcal{L}_{\mathbb{P}} &=&
       ig\text{Tr}\left[\bar{H_a}^{(\bar{Q})}\gamma^{\mu}A^{\mu}_{ab}
       \gamma_5H_b^{(\bar{Q})}\right],
     \label{lag01}\\
\mathcal{L}_{\mathcal{S}} &=&
-\frac{3}{2}g_1\varepsilon^{\mu\nu\lambda\kappa}v_{\kappa}\text{Tr}
\left[\bar{\mathcal{S}}_{\mu}A_{\nu}\mathcal{S}_{\lambda}\right],\label{lag02}\\
\mathcal{L}_{\mathcal{B}_{\bar{3}}} &=& g_{2}\text{Tr}\left[\bar{\mathcal{B}}_{\bar{3}}
\gamma_{\mu}\gamma_5A^{\mu}{\mathcal{B}}_{\bar{3}}\right],\label{lag03}
\end{eqnarray}
where $H_b^{(\bar{Q})}$ and $\mathcal{S}_{\mu}$ are defined as field operators. $H_a^{(\bar{Q})}$ is defined as $H_a^{(\bar{Q})} = [P_a^{*(\bar{Q})\mu}\gamma_{\mu}-P_a^{(\bar{Q})}\gamma_5]\frac{1-\rlap\slash v}{2}$ with the heavy pseudoscalar meson $P^{(\bar{Q})}=(\bar{D}^0,D^-, D_s^-)^T$ and heavy vector meson $P^{*(\bar{Q})}=(\bar{D}^{*0}, D^{*-}, D_s^{*-})^T$. Its conjugate field satisfies $\bar{H}_a^{(\bar{Q})}=\gamma_0H_a^{(\bar{Q})\dag}\gamma_0$. The superfield operator $\mathcal{S}_{\mu}$ is related to baryons $\mathcal{B}_6$ with $J^P=1/2^+$ and $\mathcal{B}^*_6$ with $J^P=3/2^+$ in the $6_F$ flavor representation. The expression of $\mathcal{S}_{\mu}$ reads as $\mathcal{S}_{\mu} =-\sqrt{\frac{1}{3}}(\gamma_{\mu}+v_{\mu})\gamma^5\mathcal{B}_6
+\mathcal{B}_{6\mu}^*$. Additionally, the four velocity has the form $v=(1,\bf{0})$. The axial current satisfies $A_{\mu} = \frac{1}{2}(\xi^{\dag}\partial_{\mu}\xi-\xi\partial_{\mu}\xi^{\dag})$ with $\xi=\text{exp}(i\mathbb{P}/f_{\pi})$ and the pion decay constant is taken as $f_{\pi}=132$ MeV. The concrete expressions of matrices $\mathbb{P}$, $\mathcal{B}_{\bar{3}}$, and $\mathcal{B}_{6}^{(*)}$ are
\begin{eqnarray}
\mathbb{P} &=& {\left(\begin{array}{ccc}
       \frac{\pi^0}{\sqrt{2}}+\frac{\eta}{\sqrt{6}} &\pi^+      &K^+\\
       \pi^-       &-\frac{\pi^0}{\sqrt{2}}+\frac{\eta}{\sqrt{6}}     &K^0\\
       K^-         &\bar{K}^0      &-\frac{2}{\sqrt{6}}\eta
               \end{array}\right)},\nonumber\\
\mathcal{B}_{\bar{3}} &=& {\left(\begin{array}{ccc}
         0    &\Lambda_c^+       &\Xi_c^+\\
        -\Lambda_c^+      &0     &\Xi_c^0\\
       -\Xi_c^+      &-\Xi_c^0    &0
                \end{array}\right),}\nonumber\\
\mathcal{B}_6^{(*)} &=& {\left(\begin{array}{ccc}
         \Sigma_c^{(*)++}              &\frac{1}{\sqrt{2}}\Sigma_c^{(*)+}    &\frac{1}{\sqrt{2}}\Xi_c^{(',*)+}\\
         \frac{1}{\sqrt{2}}\Sigma_c^{(*)+}      &\Sigma_c^{(*)0}       &\frac{1}{\sqrt{2}}\Xi_c^{(',*)0}\\
          \frac{1}{\sqrt{2}}\Xi_c^{(',*)+}     &\frac{1}{\sqrt{2}}\Xi_c^{(',*)0}      &\Omega_c^{(*)0}\\
                \end{array}\right).}\nonumber
\end{eqnarray}

By further expanding Eqs. (\ref{lag01}-\ref{lag03}), the concrete expressions of the effective Lagrangians adopted in our calculation can be obtained, i,e.,
\begin{eqnarray}
\mathcal{L}_{\bar{P}^*\bar{P}^*\mathbb{P}} &=&
           i\frac{2g}{f_{\pi}}v^{\alpha}\varepsilon_{\alpha\mu\nu\lambda}
           \bar{P}_{a}^{*\mu\dag}\bar{P}_{b}^{*\lambda}\partial^{\nu}\mathbb{P}_{ab},\label{hhh1}\\
\mathcal{L}_{\mathcal{B}_6\mathcal{B}_6\mathbb{P}} &=&
      i\frac{g_1}{2f_{\pi}}\varepsilon^{\mu\nu\lambda\kappa}v_{\kappa}
      \text{Tr}\left[\bar{\mathcal{B}_6}\gamma_{\mu}\gamma_{\lambda}
      \partial_{\nu}\mathbb{P}\mathcal{B}_6\right],\\
\mathcal{L}_{\mathcal{B}_6^*\mathcal{B}_6^*\mathbb{P}} &=&
      -i\frac{3g_1}{2f_{\pi}}\varepsilon^{\mu\nu\lambda\kappa}v_{\kappa}
      \text{Tr}\left[\bar{\mathcal{B}}_{6\mu}^{*}\partial_{\nu}\mathbb{P}
      \mathcal{B}_{6\lambda}^*\right],\\
\mathcal{L}_{\mathcal{B}_{\bar{3}}\mathcal{B}_{\bar{3}}\mathbb{P}} &=&
        \frac{g_{2}}{f_{\pi}}
        \text{Tr}\left[\bar{\mathcal{B}}_{\bar{3}}\gamma_{\mu}\gamma_5\mathbb{P}^{\mu}{\mathcal{B}}_{\bar{3}}\right].\label{hhh2}
\end{eqnarray}
In the above Lagrangians, there are three coupling constants, $g$, $g_{1}$, and $g_2$, to be determined. Based on the $D^*$ decay width {($\Gamma(D^{*+})=96\pm4\pm22$ keV)} \cite{Ahmed:2001xc}, the coupling constant $g$ in Eq. (\ref{lag01}) is found to be $g=0.59\pm0.01\pm 0.07$. In Eq. (\ref{lag02}), $g_1$ is fixed as $g_1= 0.94$ (Ref. \cite{Liu:2011xc})\footnote{{{In Ref. \cite{Liu:2011xc}, the coupling constant $g_1$ is related to another coupling constant $g_4$ by the relation $g_1=\frac{\sqrt{8}}{3}g_4$, where $g_4$ is a coupling constant of $\Sigma_c^*\to \Lambda_c\pi$. By the measured decay width of $\Sigma_c^*\to \Lambda_c\pi$ \cite{Lee:2014htd}, $g_4=0.999$ can be extracted, by which $g_1=0.94$ is estimated (see Ref. \cite{Liu:2011xc} for more details).}}}. The coupling constant $g_2$ in Eq. (\ref{lag03}) describes the strength of the coupling  of the pseudoscalar meson $\pi/\eta$ and charmed baryons $\mathcal{B}_{\bar{3}}$ with its light quarks in $\bar{3}_F$ representation. In the heavy quark limit, the coupling constant $g_2$ is taken as $g_2=0$ because the decay process $\mathcal{B}_{\bar{3}}\rightarrow \mathcal{B}_{\bar{3}}+ \pi/\eta$ is forbidden.

In the following, we continue to the deduction of the effective potential. The normalization relations for vector meson $P^*$, baryon $\mathcal{B}_{\bar{3}}(\mathcal{B}_6)$ with spin-1/2, and baryon $\mathcal{B}_6^{*\mu}$ with spin-3/2 are
\begin{eqnarray*}
\langle0|\bar{P}^*_{\mu}|\bar{Q}q(1)\rangle &=& \epsilon_{\mu}\sqrt{M_{\bar{P}^*}}, \\
\langle0|\mathcal{B}_{\bar{3}}|Qqq(\frac{1}{2})\rangle &=& \sqrt{2M_{\mathcal{B}_{\bar{3}}}}
        {\left(\left(1-\frac{\bf{p}^2}
        {8M_{\mathcal{B}_{\bar{3}}}^2}\right)
        \chi_{\frac{1}{2},m},
        \frac{\bf{\sigma}\cdot\bf{p}}
        {2M_{\mathcal{B}_{\bar{3}}}}
        \chi_{\frac{1}{2},m}\right)^T},\\
\langle0|\mathcal{B}_{6}^{*\mu}|Qqq(\frac{3}{2})\rangle &=&
        \sum_{m_1,m_2}C_{1/2,m_1;1,m_2}^{3/2,m_1+m_2}
        \sqrt{2M_{\mathcal{B}_{6}^*}}\nonumber\\
     &&\left(\left(1-\frac{\bf{p}^2}
        {8M_{\mathcal{B}_6^*}}\right)
        \chi_{\frac{1}{2},m_1},\frac{\bf{\sigma}
        \cdot\bf{p}}{2M_{\mathcal{B}_6^*}}
        \chi_{\frac{1}{2},m_1}\right)^T\epsilon^{\mu}_{m_2},
\end{eqnarray*}
respectively.
Here, $M_{b}$ ($b=\mathcal{B}_{\bar{3}}$,~$\mathcal{B}_{6}$,~$\bar{P}^*$) denotes the corresponding mass of vector meson $P^*$/baryon $\mathcal{B}_{\bar{3}}$/baryon $\mathcal{B}_6^{*\mu}$. The ${\bm \sigma}$ and ${\bm p}$ are the Pauli matrix and the momentum of the corresponding heavy hadron, respectively.

With the above preparation, we get the general expressions of the one-pion-exchange and one-eta-exchange effective potentials for the molecular systems considered in this work, i.e.,
\begin{eqnarray}
\mathcal{V}^{\Lambda_c\bar{D}_s^*\to \Lambda_c\bar{D}_s^*}_{\eta}({r}) &=& 0,\\
\mathcal{V}^{\Sigma_c\bar{D}_s^*\rightarrow \Sigma_c\bar{D}_s^*}_{\eta}({r}) &=&
            -\frac{1}{3}\frac{gg_1}{f_{\pi}^2}
            \mathcal{V}_1(\Lambda,m_{\eta},{r}),
            \label{sub01}\\
\mathcal{V}^{\Sigma_c^*\bar{D}_s^*\to \Sigma_c^*\bar{D}_s^*}_{\eta}({r}) &=&
            -\frac{1}{2}\frac{gg_1}{f_{\pi}^2}
            \mathcal{V}_2(\Lambda,m_{\eta},{r}),
            \label{sub02}\\
\mathcal{V}^{\Xi_c\bar{D}^*\to\Xi_c\bar{D}^*}_{\pi}({r}) &= &0,  \\
\mathcal{V}^{\Xi_c\bar{D}^*\to\Xi_c\bar{D}^*}_{\eta}({r}) &=&0,\\
\mathcal{V}^{\Xi_c'\bar{D}^*\to\Xi_c'\bar{D}^*}_{\pi}({r}) &=&\frac{1}{4}\mathcal{G}(I)\frac{gg_1}{f_{\pi}^2}
            \mathcal{V}_1(\Lambda,m_{\pi},{r}),
            \label{sub03}\\
\mathcal{V}^{\Xi_c'\bar{D}^*\to\Xi_c'\bar{D}^*}_{\eta}({r}) &=&-\frac{1}{12}\frac{gg_1}{f_{\pi}^2}
            \mathcal{V}_1(\Lambda,m_{\eta},{r}),
            \label{sub04}\\
\mathcal{V}^{\Xi_c^*\bar{D}^*\to\Xi_c^*\bar{D}^*}_{\pi}({r}) &=&\frac{3}{8}\mathcal{G}(I)\frac{gg_1}{f_{\pi}^2}
            \mathcal{V}_2(\Lambda,m_{\pi},{r}),
            \label{sub05}\\
\mathcal{V}^{\Xi_c^*\bar{D}^*\to\Xi_c^*\bar{D}^*}_{\eta}({r}) &=&-\frac{1}{8}\frac{gg_1}{f_{\pi}^2}
            \mathcal{V}_2(\Lambda,m_{\eta},{r}).
            \label{sub06}
\end{eqnarray}
Here, the subscript of $\mathcal{V}^{\Lambda_c\bar{D}_s^*\to \Lambda_c\bar{D}_s^*}_{\eta}({r})$ denotes that the exchanged meson is $\eta$. We use the same notation for the other effective potentials listed in Eqs. (\ref{sub01})-(\ref{sub06}). At first sight, the $\Lambda_c\bar{D}^*_s$ and $\Xi_c\bar{D}^*$ interactions are forbidden because of the constraint from heavy quark symmetry. For the $\Sigma_c\bar{D}_s^*$ and $\Sigma_c^*\bar{D}_s^*$ interactions, the one pion exchange is suppressed according to the OZI rule and vanishes under the symmetries considered in the current work. In the one-pion-exchange effective potentials listed in Eq. (\ref{sub03}) and Eq. (\ref{sub05}), an isospin factor $\mathcal{G}(I)$ is introduced, which is taken as $\mathcal{G}=1$ for the isovector sector with $I=1$, and $\mathcal{G}=-3$ for the isoscalar sector with $I=0$. For the convenience of the reader, two auxiliary potential functions $\mathcal{V}_1$ and $\mathcal{V}_2$ are given here, i.e.,
\begin{eqnarray}
\mathcal{V}_1(\Lambda,m,{r}) &=&
            \frac{1}{3}\left[\left(i\bf{\epsilon_1}\times
            \bf{\epsilon_3}^{\dag}\right)
            \cdot\bf{\sigma}\right]Z(\Lambda,m,{r})\nonumber\\
            &&+\frac{1}{3}{S(\hat{r},i\bf{\epsilon}_1
            \times\bf{\epsilon_3}^{\dag},\bf{\sigma})}
            T(\Lambda,m,{r}),\label{v1}\\
\mathcal{V}_2(\Lambda,m,{r}) &=&
            -\sum_{a,b,c,d}\left\langle\frac{1}{2},a;1,b
            \bigg|\frac{3}{2},a+b\right\rangle \left\langle\frac{1}{2},c;1,d\bigg|\frac{3}{2},c+d\right\rangle\nonumber\\
            &&\chi^{a\dag}_4\chi^{c}_2\left\{\frac{1}{3}
            \left(\bf{\epsilon}_1\times\bf{\epsilon}_3^{\dag}\right)
            \cdot\left(\bf{\epsilon}_2^{d}\times
            \bf{\epsilon}_4^{b\dag}\right)Z(\Lambda,m,{r})\right.\nonumber\\
            &&\left.+\frac{1}{3}S\left(\hat{r},\bf{\epsilon}_1
            \times\bf{\epsilon}_3^{\dag},
            \bf{\epsilon}_2^{d}\times\bf{\epsilon}_4^{b\dag}
            \right)T(\Lambda,m,{r})\right\},\label{v2}
\end{eqnarray}
where $S(\hat{r},\bf{x},\bf{y})= 3(\hat{r}\cdot\bf{x})(\hat{r}\cdot\bf{y})-\bf{x}\cdot\bf{y}$, and the functions $Y(\Lambda,m,r)$, $Z(\Lambda,m,r)$, and $T(\Lambda,m,r)$ have the definitions
\begin{eqnarray}
Y(\Lambda,m,{r}) &=& \frac{1}{4\pi r}(e^{-mr}-e^{-\Lambda r})-\frac{\Lambda^2-m^2}{8\pi \Lambda}e^{-\Lambda r},\label{yy}\\
Z(\Lambda,m,{r}) &=& \nabla^2Y(\Lambda,m,\bm{r}),\label{zz}\\
T(\Lambda,m,{r}) &=& r\frac{\partial}{\partial r}\frac{1}{r}\frac{\partial}{\partial r}Y(\Lambda,m,\bm{r}).\label{tt}
\end{eqnarray}

The values of the angular momentum operators in Eqs. (\ref{v1})-(\ref{v2}) sandwiched between the wave functions can be read from Table \ref{matrix}, which will be used in the calculation.

\end{multicols}
\begin{center}
\tabcaption{\label{matrix} Matrix representations for the  angular momentum operators $\langle{}^{2S'+1}L'_{J'}|\mathcal{O}_i|{}^{2S+1}L_{J}\rangle$. Here, $\mathcal{E}_1=(i\bf{\epsilon_1}\times\bf{\epsilon_3}^{\dag})\cdot\bf{\sigma}$, $\mathcal{S}_1=S(\hat{r},i\bf{\epsilon_1}\times
\bf{\epsilon_3}^{\dag},\bf{\sigma})$, $\mathcal{E}_2=\sum_{a,b;c,d}C_{1/2,a;1,b}^{3/2,m}
C_{1/2,c;1,d}^{3/2,n}\chi^{a\dag}_4\chi^{c}_2(\bf{\epsilon}_1
\times\bf{\epsilon}_3^{\dag})\cdot(\bf{\epsilon}_2^{d}
\times\bf{\epsilon}_4^{b\dag})$, and $\mathcal{S}_2=\sum_{a,b;c,d}C_{1/2,a;1,b}^{3/2,m}
C_{1/2,c;1,d}^{3/2,n}\chi^{a\dag}_4\chi^{c}_2
S(\hat{r},\bf{\epsilon}_1\times\bf{\epsilon}_3^{\dag},
\bf{\epsilon}_2^{d}\times\bf{\epsilon}_4^{b\dag})$.}
\footnotesize
\begin{tabular*}{177mm}{@{\extracolsep{\fill}}cclclclcl}
\toprule[1pt]
  {$J$}  &\quad\quad &{$\langle{}^{2S'+1}L'_{J'}|\mathcal{E}_1|{}^{2S+1}L_{J}\rangle$}     &\quad\quad
  &{$\langle{}^{2S'+1}L'_{J'}|\mathcal{S}_1|{}^{2S+1}L_{J}\rangle$}   &\quad\quad
  &{$\langle{}^{2S'+1}L'_{J'}|\mathcal{E}_2|{}^{2S+1}L_{J}\rangle$}    &\quad\quad
  &{$\langle{}^{2S'+1}L'_{J'}|\mathcal{S}_2|{}^{2S+1}L_{J}\rangle$}\\
  \midrule[1pt]
  $1/2$   &\quad &$\left(\begin{array}{cc}-2 &0\\
                           0 &1\end{array}\right)$
            &\quad &$\left(\begin{array}{cc}0 &-\sqrt{2}\\
                           -\sqrt{2} &-2\end{array}\right)$
            &\quad&$\left(\begin{array}{cc}\frac{5}{3} &0\\
                           0 &\frac{2}{3}\end{array}\right)$
           &\quad &$\left(\begin{array}{cc}0         &-\frac{7}{3\sqrt{5}}\\
                      -\frac{7}{3\sqrt{5}} &\frac{16}{15}\end{array}\right)$
           \\
  $3/2$   &\quad&$\left(\begin{array}{ccc}1  &0  &0\\
                           0  &-2 &0\\
                           0  &0  &1\end{array}\right)$
           &\quad &$\left(\begin{array}{ccc}0  &1  &2\\
                            1  &0  &-1\\
                            2  &-1 &0\end{array}\right)$
           &\quad  &$\left(\begin{array}{ccc}\frac{2}{3}  &0  &0\\
                           0  &\frac{5}{3} &0\\
                           0  &0  &\frac{2}{3}\end{array}\right)$
           &\quad  &$\left(\begin{array}{ccc}0  &\frac{7}{3\sqrt{10}}  &-\frac{16}{15}\\
                   \frac{7}{3\sqrt{10}}  &0 &-\frac{7}{3\sqrt{10}}\\
                -\frac{16}{15}  &-\frac{7}{3\sqrt{10}}  &0\end{array}\right)$
           \\
  $5/2$   &\quad&$\ldots$    &\quad &$\ldots$
           &\quad &$\left(\begin{array}{cccc}-1  &0  &0  &0\\
                           0  &\frac{5}{3} &0  &0\\
                           0  &0  &\frac{2}{3}  &0\\
                           0  &0  &0    &-1\end{array}\right)$
            &\quad&{$\left(\begin{array}{cccc}
                         0                        &\frac{2}{\sqrt{15}}       &\sqrt{\frac{3}{175}}      &-\frac{\sqrt{56}}{5}\\
                         \frac{2}{\sqrt{15}}      &0                         &\sqrt{\frac{7}{45}}       &-\sqrt{\frac{32}{105}}\\
                         \sqrt{\frac{3}{175}}     &\sqrt{\frac{7}{45}}       &-\frac{16}{21}            &-\sqrt{\frac{2}{147}}\\
                         -\frac{\sqrt{56}}{5}     &-\sqrt{\frac{32}{105}}    &-\sqrt{\frac{2}{147}}     &-\frac{4}{7}\end{array}\right)$}
                \\
\bottomrule[1pt]

\end{tabular*}%
\end{center}

\begin{multicols}{2}

With the effective potentials obtained, the bound state solutions (binding energy $E$ and corresponding root-mean-square radius $r_{RMS}$) can be obtained by solving the coupled-channel Schr$\ddot{\text{o}}$dinger equation. The corresponding kinetic terms for the systems investigated read as
\begin{eqnarray}
K_{\alpha}^{J=1/2} &=& \text{diag}\left(-\frac{\nabla^2}{2M_{\alpha}}, -\frac{\nabla_1^2}{2M_{\alpha}}\right),\\
K_{\alpha}^{J=3/2} &=& \text{diag}\left(-\frac{\nabla^2}{2M_{\alpha}}, -\frac{\nabla_1^2}{2M_2},-\frac{\nabla_1^2}{2M_{\alpha}}\right),\\
K_{\alpha}^{J=5/2} &=& \text{diag}\Bigg(-\frac{\nabla^2}{2M_{\alpha}}, -\frac{\nabla_1^2}{2M_{\alpha}},-\frac{\nabla_1^2}{2M_{\alpha}}, -\frac{\nabla_1^2}{2M_{\alpha}}\Bigg),
\end{eqnarray}
where $\nabla^2 = \frac{1}{r^2}\frac{\partial}{\partial r}r^2\frac{\partial}{\partial r}$, $\nabla_1^2 = \nabla^2-6/r^2$, and $M_{\alpha}$ is the reduced mass with the subscript $\alpha$ standing for the different systems $\Sigma_c\bar{D}_s^*$, $\Sigma_c^*\bar{D}_s^*$, $\Xi_c'\bar{D}^*$, and $\Xi_c^*\bar{D}^*$.

Recall that the $\Lambda_c\bar{D}^*_s$ and $\Xi_c\bar{D}^*$ interactions are forbidden under heavy quark symmetry and chiral symmetry. In Section~\ref{num1} and Section~\ref{num2}, we will present our results for two different types of molecular system, i.e., the $\Sigma_c^{(*)}\bar{D}_s^*$ systems, with the strange quark in the constituent meson, and the $\Xi_c^{(',*)}\bar{D}^*$ systems, with the strange quark in the constituent baryon, respectively.

\section{Numerical results}\label{sec3}

In Table~\ref{hadron}, we list the masses and quantum numbers of charmed hadrons involved in our calculation. In this work, special attention will be paid to the roles of the one-pion-exchange (OPE) potential and the one-eta-exchange (OEE) potential in forming a hadronic molecular state.

\begin{center}
\tabcaption{\label{hadron} Properties of hadrons involved in this work~\cite{Agashe:2014kda}. Here, the mass is taken as the average value, for example, $m_{\bar{D}^*}=(m_{\bar{D}^{*0}}+m_{D^{*-}})/2$.}
\footnotesize
\begin{tabular*}{85mm}{c@{\extracolsep{\fill}}cccccc}
\toprule[1pt]
  Hadrons     &$I^G(J^{P})$ &Mass (MeV)    &  Hadrons     &$I^G(J^{P})$ &Mass (MeV)\\
\midrule[1pt]
 $\bar{D}^{*} $    &$\frac{1}{2}(1^-)$    &2008.63
 &$\bar{D}_s^{*} $      &$0(1^-)$    &2112.3\\
 $\Lambda_c$      &$0(\frac{1}{2}^+)$    &2286.46
 &$\Xi_c$       &$\frac{1}{2}(\frac{1}{2}^+)$    &2469.34\\
 $\Sigma_c$     &$1(\frac{1}{2}^+)$    &2453.54
 &$\Xi_c'$     &$\frac{1}{2}(\frac{1}{2}^+)$    &2576.75\\
 $\Sigma_c^*$     &$1(\frac{3}{2}^+)$    &2518.07
 &$\Xi_c^*$     &$\frac{1}{2}(\frac{3}{2}^+)$    &2645.9\\
 $\eta$       &$0^+(0^{-})$     &547.853
 &$\pi$     &$1^-(0^-)$    &139.57\\
\bottomrule[1pt]
\end{tabular*}
\end{center}

\subsection{The $\Sigma_c\bar{D}_s^*$ and $\Sigma_c^*\bar{D}_s^*$ systems}\label{num1}

For the $\Sigma_c\bar{D}_s^*$ and $\Sigma_c^*\bar{D}_s^*$ systems, the OPE potential does not exist, and only the contribution from the OEE should be considered. In Table \ref{SigmaDs}, the corresponding bound state solutions are listed. In addition, we also present the $\Lambda$ dependence of the bound state solutions in Fig. \ref{SDs}.

\begin{center}
\tabcaption{\label{SigmaDs} {Typical values of the obtained bound state solutions (binding energy $E$ and root-mean-square radius $r_{RMS}$) for the $\Sigma_c\bar{D}_s^*$ and $\Sigma_c^*\bar{D}_s^*$ systems. $E$, $r_{RMS}$, and $\Lambda$ are in units of MeV, fm, and GeV, respectively.}}
\footnotesize
\begin{tabular*}{85mm}{c@{\extracolsep{\fill}}cccccccr}
\toprule[1pt]
   &\multicolumn{3}{c}{{$\Sigma_c\bar{D}_s^*$}}& &\multicolumn{3}{c}{$\Sigma_c^*\bar{D}_s^*$ }\\\cline{2-4}\cline{6-8}
$J$ &$\Lambda$    &$E$  &$r_{RMS}$
 &\quad\quad  &$\Lambda$     &$E$   &$r_{RMS}$  \\ \midrule[1pt]
 $1(\frac{1}{2}^-)$
    &2.88    &-0.53   &3.78
          &    &2.50       &-1.59         &2.31\\
    &2.98    &-5.52   &1.26
          &    &2.55       &-4.68         &1.36\\
    &3.08    &-15.43   &0.78
          &    &2.60       &-9.29         &0.98\\
 $1(\frac{3}{2}^-)$
    &\ldots    &\ldots   &\ldots
          &    &3.75       &-0.32        &4.52\\
    &\ldots    &\ldots   &\ldots
          &    &3.85       &-2.84        &1.78\\
    &\ldots    &\ldots   &\ldots
          &    &3.95       &-7.86        &1.09\\
 $1(\frac{5}{2}^-)$
    &\ldots    &\ldots   &\ldots
          &    &\ldots    &\ldots   &\ldots\\
\bottomrule[1pt]

\end{tabular*}
\end{center}

\end{multicols}
\begin{center}
  \includegraphics[width=5in]{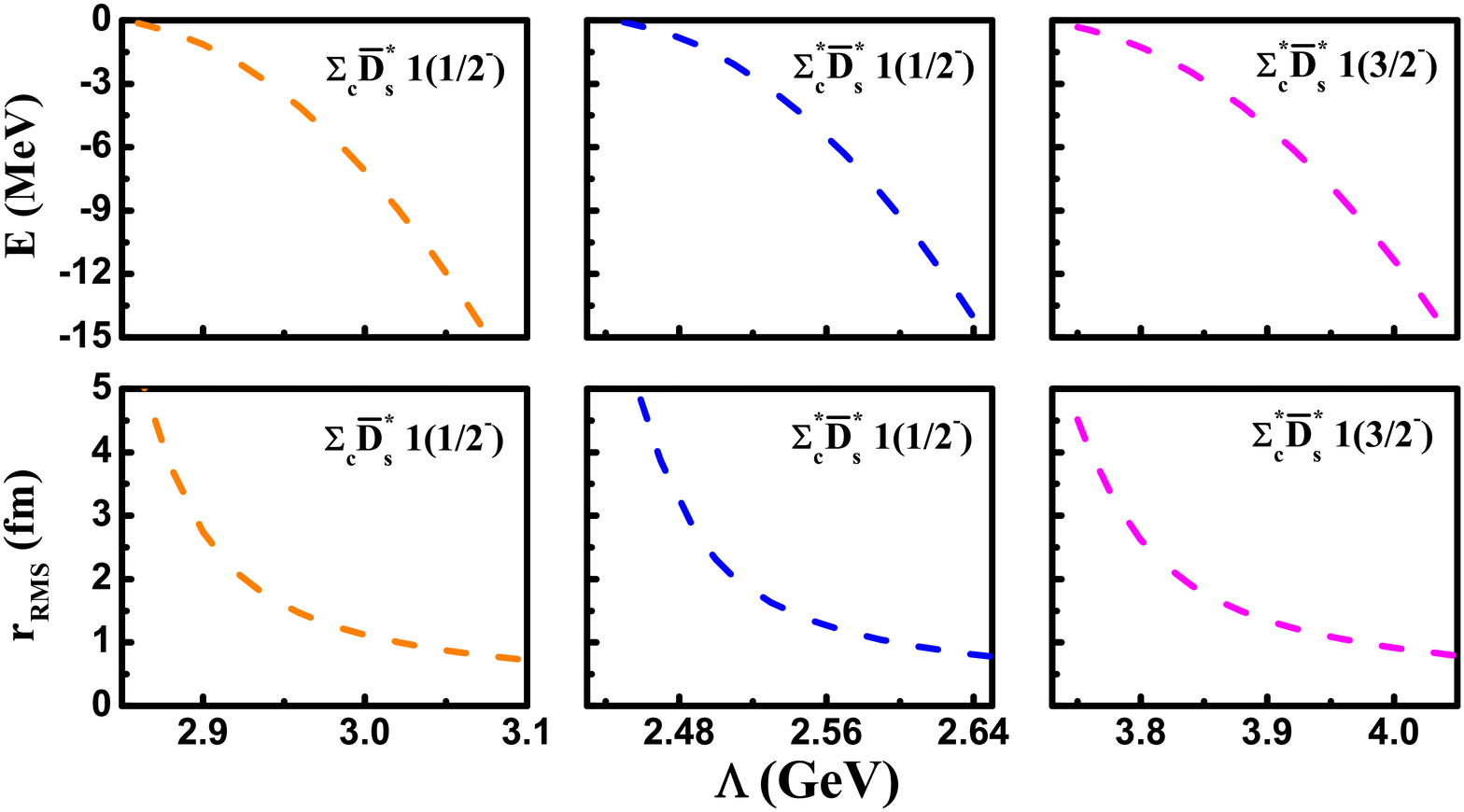}
\figcaption{\label{SDs} (color online). $\Lambda$ dependence of bound state solutions (the binding energy $E$ and the root-mean-square radius $r_{RMS}$) for the $\Sigma_c\bar{D}_s^*$ state with $1(\frac{1}{2}^-)$ and the $\Sigma_c^*\bar{D}_s^*$ states with $1(\frac{1}{2}^-)$ and $1(\frac{3}{2}^-)$.}
\end{center}
\begin{multicols}{2}

When scanning the $\Lambda$ range from 0.6 GeV to 5 GeV, we find that there exist bound state solutions for the $\Sigma_c\bar{D}_s^*$ state with quantum number $I(J^P)=1(\frac{1}{2}^-)$ and the $\Sigma_c^*\bar{D}_s^*$ states with $1(\frac{1}{2}^-)$ and $1(\frac{3}{2}^-)$. Usually, with an estimate of radius of the heavy meson about $0.5-1$ fm and by assuming the light meson to be a point particle, the cutoff should be about $1-1.5$ GeV. The corresponding $\Lambda$ values for the $\Sigma_c\bar{D}_s^*$ state with quantum number $I(J^P)=1(\frac{1}{2}^-)$ and the $\Sigma_c^*\bar{D}_s^*$ states with $1(\frac{1}{2}^-)$ and $1(\frac{3}{2}^-)$ are larger than such an estimate. This phenomenon reflects that the OEE really provides an attractive force, but it is not strong enough to produce the bound state if a cutoff about $1-1.5$ GeV is adopted strictly.

If the existence of the $\Sigma_c\bar{D}_s^*$ molecular pentaquark state with quantum number $I(J^P)=1(\frac{1}{2}^-)$ and the $\Sigma_c^*\bar{D}_s^*$ molecular pentaquark states with $1(\frac{1}{2}^-)$ and $1(\frac{3}{2}^-)$ are possible, their allowed two-body decay channels include
\begin{eqnarray}
\Sigma_c\bar{D}_s,~\Xi_c\bar{D},~\Xi_c\bar{D}^*,~\Xi_c'\bar{D},~\eta_c\Sigma,~J/\psi\Sigma,\nonumber
\end{eqnarray}
by which experiments like LHCb may search for these three strange hidden-charm molecular pentaquarks in future.

\subsection{The $\Xi_c'\bar{D}^*$ and $\Xi_c^*\bar{D}^*$ systems}\label{num2}

For the $\Xi_c'\bar{D}^*$ and $\Xi_c^*\bar{D}^*$ systems, the OPE is not suppressed, so it works with the OEE to provide the interaction force. The numerical results of bound state solutions for the $\Xi_c'\bar{D}^*$ and $\Xi_c^*\bar{D}^*$ systems are collected in Table \ref{XiD} and Table \ref{XiD2}, respectively. Here, we still scan the cutoff $\Lambda$ from 0.6 GeV to 5 GeV.

\end{multicols}
\begin{center}
\tabcaption{\label{XiD} Typical values of the bound state solutions (binding energy E and root-mean-square radius $r_{RMS}$) for the $\Xi_c'\bar{D}^*$  system. Here, $E$, $r_{RMS}$, and $\Lambda$ are in units of MeV, fm, and GeV, respectively.}
\footnotesize
\begin{tabular*}{175mm}{@{\extracolsep{\fill}}c|ccccccccc}
\toprule[1pt]
    &OEE     &\multicolumn{3}{c}{OPE}  & &\multicolumn{3}{c}{OPE\&OEE}
    \\\cline{2-9}
$I(J^P)$      &$[\Lambda, E, r_{RMS}]$ &$\Lambda$     &$E$     &$r_{RMS}$
             &\quad\quad
             &$\Lambda$     &$E$     &$r_{RMS}$
   \\ \midrule[1pt]
$0(\frac{1}{2}^-)$  &\ldots
        &1.14    &-0.20    &5.04
        &    &1.12    &-0.43    &4.18\\
        &\ldots&1.26    &-3.96    &1.62
        &  &1.22    &-4.13    &1.59\\
        &\ldots&1.38    &-13.02   &0.97
        &  &1.32    &-12.14   &1.00\\
$1(\frac{1}{2}^-)$   &\ldots
         &\ldots    &\ldots   &\dots
         & &\ldots    &\ldots   &\dots\\
         &\ldots &\ldots    &\ldots   &\dots
         &  &\ldots    &\ldots   &\dots\\
         &\ldots&\ldots    &\ldots   &\dots
         &  &\ldots    &\ldots   &\dots\\
$0(\frac{3}{2}^-)$   &\ldots
         &2.62    &-0.24    &5.17   &
         &2.48    &-0.62    &3.95\\
         &\ldots&2.90    &-3.72    &1.84    &
          &2.70    &-4.09    &1.78\\
         &\ldots&3.18    &-13.18   &1.08  &
            &2.92    &-11.83   &1.14\\
$1(\frac{3}{2}^-)$ &\ldots
         &3.84    &-0.40    &4.27   &
              &\ldots    &\ldots   &\ldots\\
         &\ldots&4.04    &-4.47    &1.48 &
          &\ldots    &\ldots   &\ldots\\
         &\ldots&4.24    &-13.66   &0.88  &
          &\ldots    &\ldots   &\ldots\\
\bottomrule[1pt]
\end{tabular*}%
\end{center}

\begin{multicols}{2}

As shown in Table \ref{XiD}, we cannot find a bound state solution for the $\Xi_c^\prime\bar{D}^*$ system if only  OEE is considered in the calculation. However, with only OPE considered, the $\Xi_c'\bar{D}^*$ states with $0(\frac{1}{2}^-)$, $0(\frac{3}{2}^-)$, and $1(\frac{3}{2}^-)$ have bound state solutions: (a) the bound state solution for the $\Xi_c'\bar{D}^*$ system with $0(\frac{1}{2}^-)$ appears when taking $\Lambda=1.1$ GeV; (b) for the $\Xi_c^\prime\bar{D}^*$ system with $0(\frac{3}{2}^-)$, we may find its bound state solution if we take $\Lambda$ to be around 3 GeV, while there is a bound state solution for the $\Xi_c^\prime\bar{D}^*$ system with $1(\frac{3}{2}^-)$ and $\Lambda\sim 4$ GeV. The corresponding cutoff is unusual and deviates from $1-1.5$ GeV. Considering this situation, we predict the existence of the $\Xi_c'\bar{D}^*$ molecular pentaquark state with $0(\frac{1}{2}^-)$ and do not recommend the $\Xi_c^\prime\bar{D}^*$ states with $1(\frac{3}{2}^-)$ and $0(\frac{3}{2}^-)$ as good candidates for a molecular pentaquark.


The bound state solutions for the $\Xi_c^\prime\bar{D}^*$ system when considering both the OPE and the OEE are presented in the columns marked  ``OPE\&OEE'' in Table~\ref{XiD}. The bound state solutions for the $\Xi_c'\bar{D}^*$ systems with $0(\frac{1}{2}^-)$ and $0(\frac{3}{2}^-)$ are obtained in the OPE\&OEE mode. Compared with the results in the OPE mode, the bound state solution for the $\Xi_c^\prime\bar{D}^*$ system with $1(\frac{3}{2}^-)$ disappears when scanning the cutoff $\Lambda$ from $0.6$ GeV to $5$ GeV, which shows that the OEE effective potential provides a repulsive force for the $\Xi_c^\prime\bar{D}^*$ system with $1(\frac{3}{2}^-)$. However, for the $\Xi_c'\bar{D}^*$ systems with $0(\frac{1}{2}^-)$ and $0(\frac{3}{2}^-)$, interaction from the OEE is weakly attractive, which makes the cutoff becomes smaller when including both OPE and OEE contributions in our calculation.

To show the contributions from the OPE and the OEE more clearly, the effective potentials for the $\Xi_c^\prime\bar{D}^*$ interaction with $0(\frac{1}{2}^-)$ when taking the cutoff $\Lambda=1.32$ GeV are shown in Fig. \ref{VXiD}. The OPE effective potentials are much larger than the OEE effective potentials and dominate the total potentials for the $V_{11}$, $V_{12}$, and $V_{22}$ cases. The curves for the OPE potentials nearly overlap with these for the total effective potentials. Besides, as indicated by the results for the bound state solution, the OEE behavior is similar to that of the OPE but with a smaller contribution to the total effective potential.

\end{multicols}
\begin{center}
\includegraphics[width=6.5in]{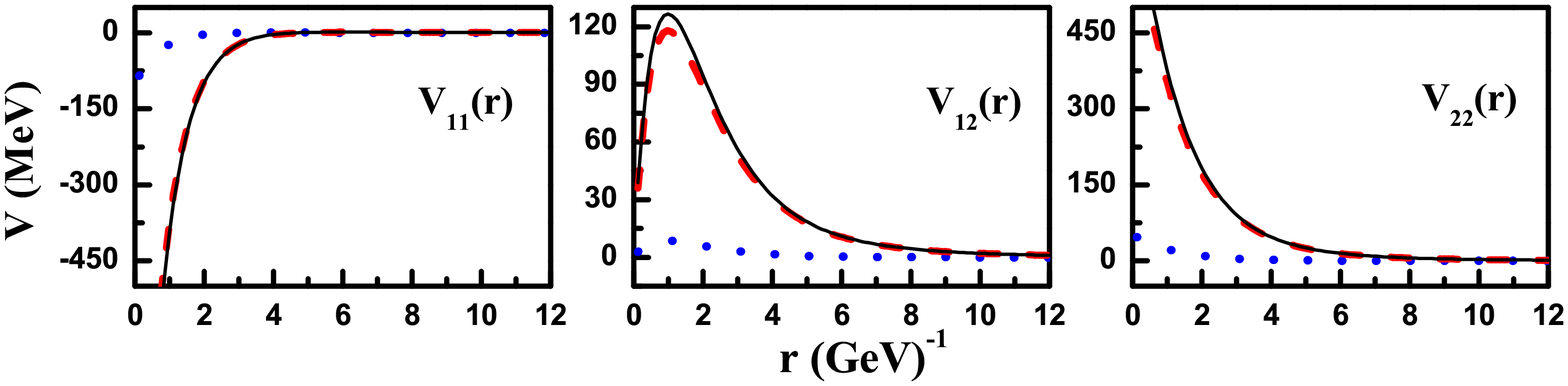}
\figcaption{\label{VXiD} (color online). Effective potentials for the $\Xi_c'\bar{D}^*$ state with $I(J^P)=0(\frac{1}{2}^-)$ at cutoff $\Lambda=1.32$ GeV. The dotted, dashed and solid lines are for the OEE , the OPE and total effective potentials, respectively. Here, $V_{11}=
  \langle{}^2\mathbb{S}_{\frac{1}{2}}|
  \mathcal{V}^{\Xi_c'\bar{D}^*\to\Xi_c'\bar{D}^*}(\bm{r})|
  {}^2\mathbb{S}_{\frac{1}{2}}\rangle$, $V_{12}=
  \langle{}^2\mathbb{S}_{\frac{1}{2}}|
  \mathcal{V}^{\Xi_c'\bar{D}^*\to\Xi_c'\bar{D}^*}(\bm{r})|
  {}^4\mathbb{D}_{\frac{1}{2}}\rangle$, and $V_{22}=
  \langle{}^4\mathbb{D}_{\frac{1}{2}}|
  \mathcal{V}^{\Xi_c'\bar{D}^*\to\Xi_c'\bar{D}^*}(\bm{r})|
  {}^4\mathbb{D}_{\frac{1}{2}}\rangle$.}
\end{center}
\begin{multicols}{2}

In Table \ref{XiD2}, we present the numerical results for the $\Xi_c^*\bar{D}^*$ system. As for the $\Xi_c'\bar{D}^*$ system, no bound state solution is obtained if only the OEE is taken into consideration. The small contribution from the OEE results in that the result with both OPE and OEE  is similar to that with the OPE only.  The cutoff to produce a $\Xi_c^*\bar{D}^*$ bound state with $0(\frac{3}{2}^-)$ is about 1.5 GeV, which shows that the $\Xi_c^*\bar{D}^*$ molecular pentaquark states with $0(\frac{3}{2}^-)$ may exist. Especially, for the $\Xi_c^*\bar{D}^*$ state with $0(\frac{1}{2}^-)$, the value of the cutoff is around $1$ GeV when the corresponding bound state solution appears. Thus, we also suggest the existence of a $\Xi_c^*\bar{D}^*$ pentaquark state with $0(\frac{1}{2}^-)$. In summary, these two $\Xi_c^*\bar{D}^*$ states with $0(\frac{1}{2}^-)$ and $0(\frac{3}{2}^-)$ are promising molecular pentaquark candidates.
If we relax the restriction on the cutoff up to 3.5 GeV, there may exist two other  molecular pentaquark state candidates, i.e., the $\Xi_c^*\bar{D}^*$ states with $0(\frac{5}{2}^-)$ and $1(\frac{5}{2}^-)$.

With the cutoff restricted to below 5 GeV, the bound state solution for the $\Xi_c^*\bar{D}^*$ system with $1(\frac{3}{2}^-)$ disappears after including the OEE contribution, while the OEE provides a repulsive force in this case.
For the $\Xi_c^*\bar{D}^*$ systems with $0(\frac{1}{2}^-)$, $0(\frac{3}{2}^-)$, and $0(\frac{5}{2}^-)$, the corresponding OEE potentials are attractive, which makes the value of the cutoff $\Lambda$ become smaller when considering both OPE and OEE contributions if reproducing the same binding energy as when only considering the OPE contribution. Different from the above three cases, the OEE contribution provides a repulsive potential for the $\Xi_c^*\bar{D}^*$ system with $1(\frac{5}{2}^-)$. Thus, we can naturally understand why the cutoff for the OPE\&OEE mode is larger than for the OPE mode, as shown in Table \ref{XiD2}.

The effective potentials for the $\Xi_c^*\bar{D}^*$ state with $0(\frac{3}{2}^-)$ are presented in Fig. \ref{VXisD}. A similar conclusion to that of the $\Xi_c'\bar{D}^*$ state with $0(\frac{3}{2}^-)$ can be reached. The OEE contribution is much smaller than the OPE contribution for all $V_{ij}$ potentials considered here, which is consistent with the observation from these results in Table \ref{XiD2}.

\end{multicols}
\begin{center}
\tabcaption{\label{XiD2} Typical values of the  bound state solutions (binding energy $E$ and root-mean-square radius $r_{RMS}$) for the $\Xi_c^*\bar{D}^*$ system. Here, $E$, $r_{RMS}$, and $\Lambda$ are in units of MeV, fm, and GeV, respectively.}
\footnotesize
\begin{tabular*}{175mm}{@{\extracolsep{\fill}}c|cccccccccccccccccc}
\toprule[1pt]

    &OEE     &\multicolumn{3}{c}{OPE}      &\quad\quad  &\multicolumn{3}{c}{OPE\&OEE}
    \\\cline{2-9}
$I(J^P)$      &$[\Lambda, E, r_{RMS}]$  &$\Lambda$     &$E$     &$r_{RMS}$  &\quad\quad
             &$\Lambda$     &$E$     &$r_{RMS}$\\ \midrule[1pt]
$0(\frac{1}{2}^-)$
        &\ldots  &0.95      &-0.14        &5.39   &   &0.95      &-0.45        &4.10\\
        &\ldots  &1.05      &-3.17        &1.80   &   &1.05      &-4.73        &1.51\\
        &\ldots  &1.15      &-10.46       &1.08   &   &1.15      &-14.27       &0.95\\

$1(\frac{1}{2}^-)$
        &\ldots  &\ldots    &\ldots    &\ldots   &   &\ldots    &\ldots    &\ldots\\

$0(\frac{3}{2}^-)$
        &\ldots  &1.55      &-0.23        &5.02   &   &1.50      &-0.48        &4.08\\
        &\ldots  &1.70      &-3.61        &1.73   &   &1.65      &-5.06        &1.50\\
        &\ldots  &1.85      &-11.86       &1.04   &   &1.80      &-15.64       &0.92\\

$1(\frac{3}{2}^-)$
        &\ldots    &4.25      &-0.44        &4.09    &  &\ldots    &\ldots   &\ldots\\
        &\ldots    &4.40      &-3.11        &1.73    &  &\ldots    &\ldots   &\ldots\\
        &\ldots    &4.55      &-8.47        &1.08    &  &\ldots    &\ldots   &\ldots\\

$0(\frac{5}{2}^-)$
        &\ldots    &2.35      &-0.24        &5.20    &  &2.20      &-0.37        &4.70\\
        &\ldots    &2.60      &-3.48        &1.93    &  &2.40      &-3.22        &2.00\\
        &\ldots    &2.85      &-12.30       &1.15    &  &2.60      &-10.07       &1.26\\

$1(\frac{5}{2}^-)$
        &\ldots    &3.15      &-0.62        &3.62    &   &4.30      &-0.26        &4.79\\
        &\ldots    &3.30      &-4.03        &1.54    &   &4.50      &-3.22        &1.70\\
        &\ldots    &3.45      &-10.84       &0.98    &   &4.70      &-10.03       &1.00\\
\bottomrule[1pt]
\end{tabular*}%
\end{center}

\begin{center}
  \includegraphics[width=6.4in]{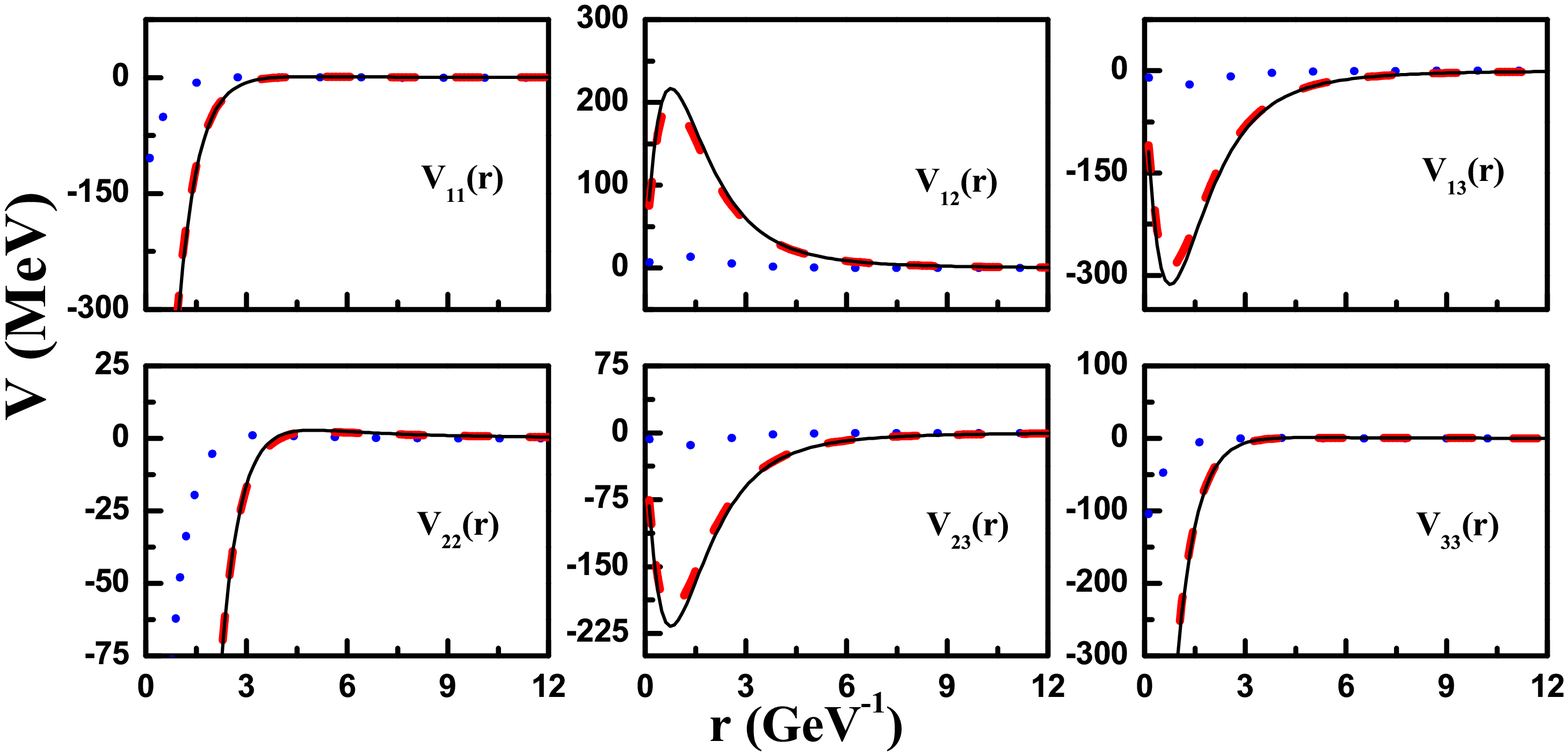}
\figcaption{\label{VXisD} (color online). Effective potentials for the $\Xi_c^*\bar{D}^*$ state with $I(J^P)=0(\frac{3}{2}^-)$ at cutoff $\Lambda=1.80$ GeV. The dotted, dashed and solid lines are for the OEE, the OPE and  total effective potentials, respectively. Here, we define $V_{ij}=
  \langle i|
  \mathcal{V}^{\Xi_c^*\bar{D}^*\to\Xi_c^*\bar{D}^*}(\bm{r})|
  j\rangle$ with $|i\rangle=|
  {}^6\mathbb{S}_{\frac{5}{2}}\rangle$, $|{}^2\mathbb{D}_{\frac{5}{2}}\rangle$, $|
  {}^4\mathbb{D}_{\frac{5}{2}}\rangle$, and $|
  {}^6\mathbb{D}_{\frac{5}{2}}\rangle$ for $i=1$, 2, 3, and 4, respectively.}
\end{center}
\begin{multicols}{2}

Additionally, we also provide the allowed two-body decay channels for the $\Xi_c^{(',*)}\bar{D}^*$ molecular pentaquarks with different quantum numbers in Table \ref{decay}, which may be useful for the further experimental study of these molecular pentaquarks.

\begin{center}
\tabcaption{\label{decay} Allowed decay channels for $\Xi_c'\bar{D}^*$ and $\Xi_c^*\bar{D}^*$ with different quantum numbers.}
\footnotesize
\begin{tabular*}{86mm}{c@{\extracolsep{\fill}}rrrrrrr}
\toprule
   &\multicolumn{2}{c}{$\Xi_c'\bar{D}^*\left[I(J^P)\right]$}  &\quad\quad  &\multicolumn{3}{c}{$\Xi_c^*\bar{D}^*\left[I(J^P)\right]$}
   \\\cline{2-3}\cline{5-7}
Channels     &$0(\frac{1}{2}^-)$\quad    &\quad$0(\frac{3}{2}^-)$\quad
   &&$0(\frac{1}{2}^-)$\quad     &\quad$0(\frac{3}{2}^-)$\quad    &\quad$0(\frac{5}{2}^-)$\quad\\
\midrule[1pt]
$\Lambda_c\bar{D}_s$    &$\checkmark$    &$\checkmark$    &&$\checkmark$     &$\checkmark$  &$\checkmark$\\
$\Lambda_c\bar{D}_s^*$  &$\checkmark$    &$\checkmark$    &&$\checkmark$     &$\checkmark$  &$\checkmark$\\
$\Xi_c\bar{D}$    &$\checkmark$    &$\checkmark$    &&$\checkmark$   &$\checkmark$  &$\checkmark$  \\
$\Xi_c\bar{D}^*$  &$\checkmark$    &$\checkmark$    &&$\checkmark$   &$\checkmark$  &$\checkmark$  \\
$\Xi_c'\bar{D}$   &$\checkmark$    &$\checkmark$    &&$\checkmark$   &$\checkmark$  &$\checkmark$  \\
$\Xi_c'\bar{D}^*$ &    &    &&$\checkmark$   &$\checkmark$  &$\checkmark$  \\
$\Xi_c^*\bar{D}$  &$\checkmark$    &$\checkmark$    &&$\checkmark$   &$\checkmark$  &$\checkmark$  \\
$\eta_c\Lambda$   &$\checkmark$    &$\checkmark$    &&$\checkmark$     &$\checkmark$  &$\checkmark$\\
$J/\psi\Lambda$   &$\checkmark$    &$\checkmark$    &&$\checkmark$     &$\checkmark$  &$\checkmark$\\
$\Xi_{cc}(\frac{1}{2}^+)\bar{K}$   &$\checkmark$    &$\checkmark$    &&$\checkmark$   &$\checkmark$  &$\checkmark$  \\
$\Xi_{cc}(\frac{1}{2}^+)\bar{K}^*$   &$\checkmark$    &$\checkmark$    &&$\checkmark$   &$\checkmark$  &$\checkmark$  \\
$\Omega_{cc}(\frac{1}{2}^+)\eta$   &$\checkmark$    &$\checkmark$    &&$\checkmark$     &$\checkmark$  &$\checkmark$\\
$\Omega_{cc}(\frac{1}{2}^+)\omega$  &$\checkmark$    &$\checkmark$    &&$\checkmark$     &$\checkmark$  &$\checkmark$\\
\bottomrule
\end{tabular*}
\end{center}

\section{Summary}\label{sec4}

Searching for exotic hadronic states is a research field full of challenges and opportunities. With recent experimental progress, more and more novel phenomena have been revealed in experiments, which has stimulated theorists' extensive interest in studying exotic states. Interested readers may read about the relevant progress in the review papers in Refs. \cite{Chen:2016qju,Chen:2016spr}.

In 2015, the observation of two $P_c$ states at LHCb \cite{Aaij:2015tga} inspired many new investigations of hidden-charm pentaquarks, and molecular assignments to the two $P_c$ states  are a popular explanation \cite{Chen:2015loa,Chen:2016heh,Chen:2015moa,Roca:2015dva,Mironov:2015ica,He:2015cea,Meissner:2015mza,
Burns:2015dwa,Huang:2015uda}. In this situation, we have reason to believe that there should exist partners of the two $P_c$ states. Thus, we need to perform dynamical studies relevant to their partners. In this work, we mainly focus on the $\Lambda_c\bar{D}_s^*/\Sigma_c^{(*)}\bar{D}_s^*/\Xi^{(\prime,*)}_c\bar{D}^*$ interactions and predict the existence of some strange hidden-charm molecular pentaquarks, as described in Section~\ref{sec3}.

Our numerical results show that the most promising strange hidden-charm molecular pentaquarks are a $\Xi_c'\bar{D}^*$ state with $0(\frac{1}{2}^-)$ and the $\Xi_c^*\bar{D}^*$ states with $0(\frac{1}{2}^-)$ and $0(\frac{3}{2}^-)$. Thus, we strongly suggest that experimentalists search for these.

In summary, the two observed $P_c$ states from LHCb have opened fascinating new avenues of research. In future, theorists and experimentalists should make more effort to study hidden-charm pentaquarks, especially the partners of $P_c(4380)$ and $P_c(4450)$. As we face this research field full of challenges, more opportunities are waiting for us.

\acknowledgments{We thank T. Burns for useful discussions.}

\end{multicols}

\vspace{10mm}

\begin{multicols}{2}

\end{multicols}

\clearpage

\end{CJK*}
\end{document}